\documentclass[12pt]{article}
\begin{document}
\vspace*{-.6in}
\thispagestyle{empty}
\begin{flushright}
CALT-68-2184\\
hep-th/9807135
\end{flushright}
\baselineskip = 20pt

\vspace{.5in}
{\Large
\begin{center}
From Superstrings to M Theory\footnote{Work
supported in part by the U.S. Dept. of Energy under Grant No.
DE-FG03-92-ER40701.}
\end{center}}

\begin{center}
John H. Schwarz\\
\emph{California Institute of Technology, Pasadena, CA  91125, USA}
\end{center}
\vspace{1in}

\begin{center}
\textbf{Abstract}
\end{center}
\begin{quotation}
\noindent  
In the strong coupling limit type IIA 
superstring theory develops an eleventh
dimension that is not apparent in perturbation theory. This suggests the
existence of a consistent 11d quantum theory, called M theory, which is
approximated by 11d supergravity at low energies. In this review we 
describe some of the evidence for this picture and some of its implications.
\end{quotation}

\vfil
\centerline{\it To be published in a special issue of Physics Reports}
\centerline{\it in memory of Richard Slansky}

\newpage

\pagenumbering{arabic}

\section{Introduction}

Superstring theory is currently undergoing a 
period of rapid development in which
important advances in understanding are being 
achieved.  The purpose of this
review is to describe a portion of this story 
to physicists who are not already
experts in this field.\footnote{For a more 
detailed review see ref.~\cite{sen98}.}
The focus will be on explaining why there can be an
eleven-dimensional vacuum, even though there are
only ten dimensions in perturbative
superstring theory.  The nonperturbative
extension of superstring theory that allows for an
eleventh dimension has been named {\em M theory}.  
The letter M is intended to
be flexible in its interpretation.  It could stand 
for {\em magic,} {\em mystery,}
or {\em meta} to reflect our current state of 
incomplete understanding.  Those
who think that two-dimensional supermembranes 
(the M2-brane) are fundamental
may regard M as standing for {\em membrane.}  
An approach called {\em Matrix
theory}  is another possibility.  And, of course, 
some view M theory as the
{\em mother} of all theories.

Superstring theory first achieved widespread acceptance during the {\em first
superstring revolution} in 1984-85.  There were three main developments at
this time.  The first was the discovery of an anomaly cancellation mechanism \cite{green84},
which showed that supersymmetric gauge theories can be consistent in ten dimensions
provided they are coupled to supergravity (as in type I superstring theory)
and the gauge group is either SO(32) or $E_8 \times E_8$.\footnote{A discussion 
with Richard Slansky helped to convince us that $E_8 \times E_8$ would work.}
Any other group
necessarily would give uncancelled gauge anomalies and hence inconsistency at
the quantum level.  The second development was the discovery of two new
superstring theories---called {\em heterotic} string theories---with precisely these
gauge groups \cite{gross84}.  The third development was the realization that the $E_8 \times
E_8$ heterotic string theory admits solutions in which six of the space dimensions
form a Calabi--Yau space, and that this results in a 4d effective theory at low
energies with many qualitatively realistic features \cite{candelas85}.  
Unfortunately, there are
very many Calabi--Yau spaces and a whole range of additional choices
that can be made (orbifolds,
Wilson loops, etc.).  Thus there is an enormous variety of possibilities, none of
which stands out as particularly special.

In any case, after the first superstring revolution subsided, we had five
distinct superstring theories with consistent weak coupling perturbation
expansions, each in ten dimensions.  Three of them, the {\em type I} theory and
the two heterotic theories, have ${\cal N} = 1$ supersymmetry in the ten-dimensional
sense.  Since the minimal 10d spinor is simultaneously Majorana and Weyl, this
corresponds to 16 conserved supercharges.  The other two theories, called {\em type
IIA} and {\em type IIB}, have ${\cal N} = 2$ supersymmetry (32 supercharges)~\cite{green82}.  
In the IIA case the two spinors have opposite handedness so that the spectrum is
left-right symmetric (nonchiral).  In the IIB case the two spinors have the
same handedness and the spectrum is chiral.

The understanding of these five superstring theories was developed in the
ensuing years.  In each case it became clear, and was largely proved, that
there are consistent perturbation expansions of on-shell scattering amplitudes.
In four of the five cases (heterotic and type II) the fundamental strings are
oriented and unbreakable.  As a result, these theories have particularly simple
perturbation expansions.  Specifically, there is a unique Feynman diagram at
each order of the loop expansion.  The Feynman diagrams depict string world sheets, and
therefore they are two-dimensional surfaces.  For these four theories the
unique $L$-loop diagram is a closed orientable
genus-$L$ Riemann surface, which can be visualized
as a sphere with $L$ handles.  External (incoming or outgoing) particles are
represented by $N$ points (or ``punctures'') on the Riemann surface.  A
given diagram represents a well-defined integral of dimension $6L + 2N - 6$.  This
integral has no ultraviolet divergences, even though the spectrum contains
states of arbitrarily high spin (including a massless graviton).  From the
viewpoint of point-particle contributions, string and supersymmetry properties
are responsible for incredible cancellations.  Type I superstrings are
unoriented and breakable.  As a result, the perturbation expansion is more
complicated for this theory, and the various world-sheet diagrams at a given order 
(determined by the Euler number) have to be
combined properly to cancel divergences and anomalies ~\cite{green85}.

An important discovery that was made between the two superstring revolutions
is called {\em T duality} \cite{giveon94}.  This is a property of string theories that can be
understood within the context of perturbation theory.  (The discoveries
associated with the {\em second superstring revolution} are mostly nonperturbative.)  T
duality shows that spacetime geometry, as probed by strings, has some
surprising properties (sometimes referred to
as {\em quantum geometry}).  The basic idea can be illustrated by the
simplest example.  This entails considering one spatial dimension to form a
circle (denoted $S^1$).  Then the ten-dimensional geometry is $R^9 \times S^1$.
 T duality identifies this string compactification with one of a second string
theory also on $R^9 \times S^1$.  However, if the radii of the circles in the
two cases are denoted $R_1$ and $R_2$, then
\begin{equation}
R_1 R_2 = \alpha'. \label{Tdual}
\end{equation}
Here $\alpha' = \ell_s^2$ is the universal Regge slope parameter, and $\ell_s$
is the fundamental string length scale (for both string theories).  The tension
of a fundamental string is given by
\begin{equation}
T = 2\pi m_s^2 = \frac{1}{2\pi\alpha'},
\end{equation}
where we have introduced a fundamental string mass scale 
\begin{equation}
m_s = (2\pi\ell_s)^{-1}.
\end{equation}

Note that T duality implies that shrinking the circle to zero in one theory
corresponds to decompactification of the dual theory.  Compactification on a
circle of radius $R$ implies that momenta in that direction are quantized, $p =
n/R$.  (These are called {\em Kaluza--Klein excitations}.)  
These momenta appear as masses for states that are
massless from the higher-dimensional viewpoint. String theories
also have a second class of excitations, called {\em winding modes}.  
Namely, a string wound $m$
times around the circle has energy $E = 2\pi R \cdot m \cdot T = mR/\alpha'$.
Equation~(\ref{Tdual}) shows that 
the winding modes and Kaluza--Klein excitations are
interchanged under T duality.

What does T duality imply for our five superstring theories?  The IIA and IIB
theories are T dual \cite{ginsparg87}.  
So compactifying the nonchiral IIA theory on a circle of
radius $R$ and letting $R \rightarrow 0$ gives the chiral IIB theory in ten
dimensions!  This means, in particular, that they should not be regarded as
distinct theories.  The radius $R$ is actually a $vev$ of a scalar field, which arises as
an internal component of the 10d metric tensor.  Thus the type IIA and type IIB
theories in 10d are two limiting points in a continuous moduli space of quantum vacua.
The two heterotic theories are also T
dual, though there are technical details involving Wilson loops, which we will not
explain here.  T duality applied to the type I theory gives a dual description,
which is sometimes called I'.  The names IA and IB have also been introduced
by some authors.

For the remainder of this paper, we will restrict attention to theories with
maximal supersymmetry (32 conserved supercharges).  This is sufficient to
describe the basic ideas of M theory.  Of course, it suppresses many
fascinating and important issues and discoveries.  In this way we will keep the
presentation from becoming too long or too technical.  The main focus will be
to ask what happens when we go beyond perturbation theory and allow the
coupling strength to become large in the type II theories.  The answer in the
IIA case, as we will see, is that another spatial dimension appears.

\section{M Theory}

In the 1970s and 1980s various supersymmetry and 
supergravity theories were constructed. (See~\cite{salam}, for example.)
In particular, supersymmetry representation theory showed that ten is the largest
spacetime dimension in which there can be a matter theory (with
spins $\leq 1$) in which supersymmetry is realized linearly.  
A realization of this is 10d super Yang--Mills
theory, which has 16 supercharges \cite{brink77}.  
This is a pretty ({\it i.e.}, very symmetrical) classical field theory, but
at the quantum level it is both nonrenormalizable and 
anomalous for any nonabelian gauge group.  However, as
we indicated earlier, both problems can be overcome for suitable gauge groups
(SO(32) or $E_8 \times E_8$) when the Yang--Mills theory 
is embedded in a type I or heterotic string theory.

The largest possible spacetime dimension for a supergravity theory (with spins $\leq 2$),
on the other hand, is eleven.  Eleven-dimensional supergravity, which has 32 conserved
supercharges, was constructed 20 years ago \cite{cremmer78a}.  
It has three kinds of fields---the
graviton field (with 44 polarizations), the gravitino field (with 128
polarizations), and a three-index gauge field $C_{\mu\nu\rho}$ (with 84
polarizations).  These massless particles are referred to
collectively as the {\em supergraviton}. 
11d supergravity is also a pretty classical field theory, which has attracted
a lot of attention over the years.  It is not chiral, and therefore not subject
to anomaly problems.\footnote{Unless the spacetime has boundaries.  The
anomaly associated to a 10d boundary can be cancelled by introducing $E_8$
supersymmetric gauge theory on the boundary \cite{horava95}.}  
It is also nonrenormalizable, and thus it cannot be a fundamental theory. 
Though it is difficult to demonstrate explicitly that it is not
finite as a result of ``miraculous'' cancellations, we now know that this
is not the case.   However, we now
believe that it is a low-energy effective description of M theory, which is a
well-defined quantum theory~\cite{witten95a}.  
This means, in particular, that higher dimension
terms in the effective action for the supergravity fields have uniquely determined
coefficients within the M theory setting, even though they are formally infinite 
(and hence undetermined) within the supergravity context.

Intriguing connections between type IIA string theory and 11d supergravity have
been known for a long time.  If one 
carries out {\em dimensional reduction} of 11d supergravity
to 10d, one gets type IIA supergravity~\cite{campbell84}.  
Dimensional reduction can be viewed as
a compactification on circle in which one drops all the Kaluza--Klein
excitations.  It is easy to show that this does not break any of the
supersymmetries.

The field equations of 11d supergravity admit a solution that describes a
supermembrane.  In other words, this solution has the property that the energy
density is concentrated on a two-dimensional surface.  A 3d world-volume
description of the dynamics of
this supermembrane, quite analogous to the 2d world volume
actions of superstrings, has been constructed~\cite{bergshoeff87}.  
The authors suggested that a
consistent 11d quantum theory might be defined in terms of this membrane, in analogy
to string theories in ten dimensions.\footnote{Most experts now 
believe that M theory cannot be defined
as a supermembrane theory.}  Another striking result was the discovery of
double dimensional reduction~\cite{duff87}.  
This is a dimensional reduction in which one
compactifies on a circle, wraps one dimension of the membrane around the circle
and drops all Kaluza--Klein excitations for both the spacetime theory and the
world-volume theory.  The remarkable fact is that this gives the (previously
known) type IIA superstring world-volume action~\cite{green84b}.

For many years these facts remained unexplained curiosities until they were
reconsidered by Townsend~\cite{townsend95a} and by Witten~\cite{witten95a}.  
The conclusion is that type IIA
superstring theory really does have a circular 11th dimension in addition to
the previously
known ten spacetime dimensions.  This fact was not recognized earlier
because the appearance of the 11th dimension is a nonperturbative phenomenon,
not visible in perturbation theory.

To explain the relation between M theory and type IIA string theory, a good
approach is to identify the parameters that characterize each of them and to
explain how they are related.  Eleven-dimensional supergravity (and hence M
theory, too) has no dimensionless parameters.  As we have seen, there are no
massless scalar fields, whose vevs could give parameters.  The only
parameter  is the 11d Newton constant, which raised to a suitable power 
($-1/9$), gives the 11d Planck mass $m_p$.  
When M theory is compactified on a
circle (so that the spacetime geometry is $R^{10} \times S^1$) another
parameter is the radius $R$ of the circle.

Now consider the parameters of type IIA superstring theory.  They are the
string mass scale $m_s$, introduced earlier, and the dimensionless string
coupling constant $g_s$.  An important fact about all five superstring theories
is that the coupling constant is not an arbitrary parameter.  Rather, it is a
dynamically determined vev of a scalar field, the {\em dilaton,} which is a
supersymmetry partner of the graviton.  With the usual conventions, one has
$g_s = \langle e^\phi\rangle$.

We can identify compactified M theory with type IIA superstring theory by
making the following correspondences:
\begin{equation}\label{M1}
m_s^2 = 2\pi R m_p^3
\end{equation}
\begin{equation}\label{M2}
g_s = 2\pi Rm_s.
\end{equation}
Using these one can derive other equivalent relations, such as
\begin{equation}
g_s = (2\pi Rm_p)^{3/2}
\end{equation}
\begin{equation}
m_s = g_s^{1/3} m_p.
\end{equation}
The latter implies that the 11d Planck length is shorter than the string length
scale at weak coupling by a factor of $(g_s)^{1/3}$.

Conventional string perturbation theory is an expansion in powers of $g_s$ at
fixed $m_s$.  Equation~(\ref{M2}) shows that this is equivalent to an expansion
about $R=0$.  In particular, the strong coupling limit of type IIA superstring
theory corresponds to decompactification of the eleventh dimension, so in a sense M theory
is type IIA string theory at infinite coupling.\footnote{The $E_8 \times E_8$ heterotic
string theory is also eleven-dimensional at strong coupling \cite{horava95}.}
This explains why the eleventh dimension was not discovered in
studies of string perturbation theory.

These relations encode some interesting facts.  The fact relevant to eq.~(\ref{M1})
concerns the interpretation of the fundamental type IIA string.
Earlier we discussed the old notion of double dimensional reduction, which
allowed one to derive the IIA superstring world-sheet action from the 
11d supermembrane (or M2-brane)
world-volume action.  Now we can make a stronger statement:  The fundamental
IIA string actually {\em is} an M2-brane of M theory with one of its dimensions
wrapped around the circular spatial dimension.   No truncation to zero modes is
required. Denoting the string and membrane tensions (energy
per unit volume) by $T_{F1}$ and $T_{M2}$, one deduces that
\begin{equation}
T_{F1} = 2\pi R \, T_{M2}.
\end{equation}
However, $T_{F1} = 2\pi m_s^2$ and $T_{M2} = 2\pi m_p^3$.  Combining these
relations gives eq.~(\ref{M1}).

Type II superstring theories contain a variety of $p$-brane solutions that preserve
half of the 32 supersymmetries. These are solutions in which
the energy  is concentrated on a $p$-dimensional spatial
hypersurface. (The  world volume has $p+1$ dimensions.)
The corresponding solutions of  supergravity
theories were constructed by Horowitz and Strominger~\cite{horowitz91}.
A large class of these $p$-brane excitations are called
{\em D-branes} (or D$p$-branes when we want to specify the dimension),
 whose tensions are given by~\cite{polchinski95}
\begin{equation}
T_{Dp} = 2\pi {m_s^{p+1}}/{g_s}.
\end{equation}
This dependence on the coupling constant is one of the characteristic features
of a D-brane.  It is to be contrasted with the more familiar 
$g^{-2}$ dependence of soliton masses
(e.g., the 't Hooft--Polyakov monopole).
Another characteristic feature of D-branes
is that they carry a charge that couples to a gauge field
in the RR sector of the theory. (Such fields can be described as bispinors.)
The particular RR gauge fields that 
occur imply that
even values of $p$ occur in the IIA theory and odd values in the IIB theory.

In particular, the D2-brane of the type IIA theory
corresponds to our friend the supermembrane of M theory, but now
in a background geometry in which one of the transverse dimensions is a circle.
The tensions check, because (using eqs.~(\ref{M1}) and~(\ref{M2}))
\begin{equation}
T_{D2} = 2\pi {m_s^3}/{g_s} = 2\pi m_p^3 = T_{M2}.
\end{equation}
The mass of the first Kaluza--Klein excitation 
of the 11d supergraviton is $1/R$.  Using eq.~(\ref{M2}),
we see that this can be identified with the D0-brane.
More identifications of this type arise when we consider the magnetic dual of
the M theory supermembrane.  This turns out to be a five-brane, called the
M5-brane.\footnote{In general, the magnetic dual of a $p$-brane in $d$
dimensions is a $(d - p - 4)$-brane.}  Its tension is $T_{M5} = 2\pi m_p^6$.
Wrapping one of its dimensions around the circle gives the D4-brane, with
tension
\begin{equation}
T_{D4} = 2\pi R \,T_{M5} = 2\pi m_s^5/g_s.
\end{equation}
If, on the other hand, the M5-frame is not wrapped around the circle, one
obtains the NS5-brane of the IIA theory with tension
\begin{equation}
T_{NS5} = T_{M5} = 2\pi m_s^6/g_s^2.
\end{equation}
This 5-brane, which is the magnetic dual of the fundamental IIA string, exhibits
the conventional $g^{-2}$ solitonic dependence.

To summarize, type IIA superstring theory is M theory compactified on a circle
of radius $R=g_s \ell_s$.
 M theory is believed to be a well-defined quantum theory in 11d, which is
approximated at low energy by 11d supergravity.  Its excitations are the
massless supergraviton, the M2-brane, and the M5-brane.  These account both for
the (perturbative) fundamental string of the IIA theory and for many of its
nonperturbative excitations.  The identities that we have presented here are exact,
because they are protected by supersymmetry.

\section{Type IIB Superstring Theory}

In the previous section we discussed type IIA superstring theory and its
relationship to eleven-dimensional M theory.  In this section we consider type
IIB superstring theory, which is the other maximally supersymmetric
string theory with
32 conserved supercharges.  It is also 10-dimensional, but unlike the IIA
theory its two supercharges have the same handedness.  Since the spectrum
contains massless chiral fields, one should check whether there are anomalies
that break the gauge invariances---general coordinate invariance, local Lorentz
invariance, and local supersymmetry.  In fact, the UV finiteness of the 
string theory Feynman
diagrams (and associated {\em modular invariance}) ensures that all anomalies must
cancel.  This was verified also from a field theory viewpoint~\cite{alvarez83}.

The low-energy effective theory that approximates type
IIB superstring theory is type IIB supergravity~\cite{green82,schwarz83}, 
just as 11d supergravity approximates M theory.  In each case the
supergravity theory is only well-defined as a classical field theory, but still
it can teach us a lot.  For example, it can be used to construct $p$-brane
solutions and compute their tensions.  Even though such solutions themselves are
only approximate, supersymmetry considerations ensure that their tensions,
which are related to the kinds of charges they carry, are exact.

Another significant fact about type IIB supergravity is that it possesses a
global $SL(2,R)$ symmetry.  It is instructive to consider the bosonic spectrum
and its $SL(2,R)$ transformation properties.  There are two scalar fields---the
dilation $\phi$ and an {\em axion} $\chi$, which are conveniently combined in a
complex field
\begin{equation}
\rho = \chi + ie^{-\phi}.
\end{equation}
The $SL(2,R)$ symmetry transforms this field nonlinearly:
\begin{equation}
\rho \rightarrow \frac{a\rho + b}{c\rho + d},
\end{equation}
 where $a,b,c,d$ are real numbers satisfying $ad - bc = 1$.  However, in the
quantum string theory this symmetry is broken to the discrete subgroup
$SL(2,Z)$~\cite{hull94}, which means that $a,b,c,d$ are restricted to be integers.  Defining
the vev of the $\rho$ field to be
\begin{equation}
\langle \rho \rangle = \frac{\theta}{2\pi} + \frac{i}{g_s},
\end{equation}
the $SL(2,Z)$ symmetry transformation $\rho \rightarrow \rho + 1$ implies that $\theta$
is an angular coordinate.  More significantly, in the special case $\theta =
0$, the symmetry transformation $\rho \rightarrow - 1/\rho$ takes $g_s
\rightarrow 1/g_s$.  This symmetry, called {\em S duality}, implies that the theory
with coupling constant $g_s$ is equivalent to coupling constant $1/g_s$, so
that the weak coupling expansion and the strong coupling expansion are
identical!

The bosonic spectrum also contains a pair of two-form potentials
$B_{\mu\nu}^{(1)}$ and $B_{\mu\nu}^{(2)}$, which transform as a doublet under
$SL(2,R)$ or $SL(2,Z)$.  In particular, the S duality transformation $\rho
\rightarrow - 1/\rho$ interchanges them.  The remaining bosonic fields are the
graviton and a four-form potential $C_{\mu\nu\rho\lambda}$, with a self-dual
field strength.  They are invariant under $SL(2)$.

In the introductory section we indicated that the type IIA and type IIB
superstring theories are T dual, meaning that if they are compactified on
circles of radii $R_A$ and $R_B$ one obtains equivalent theories for the
identification $R_AR_B = \ell_s^2$.  Moreover, in sect. 2 we saw that the
type IIA theory is actually M theory compactified on a circle.  The latter fact
encodes nonperturbative information.  It turns out to be very useful to combine
these two facts and to consider the duality between M theory compactified on a
torus $(R^9 \times T^2)$ and type IIB superstring theory compactified on a
circle $(R^9 \times S^1)$.

Recall that a torus can be described as the complex plane modded out by the
equivalence relations $z \sim z + w_1$ and $z \sim z + w_2$.  Up to conformal
equivalence, the periods can be taken to be $1$ and $\tau$, with Im $\tau >
0$.  However, in this characterization $\tau$ and $\tau' = (a\tau + b)/(c\tau +
d)$, where $a,b,c,d$ are integers satisfying $ad - bc = 1$, describe equivalent
tori.  Thus a torus is characterized by a modular parameter $\tau$ and an
$SL(2,Z)$ modular group.  The natural, and correct, conjecture at this point is
that one should identify the modular parameter $\tau$ of the M theory torus
with the parameter $\rho$ that characterizes the 
type IIB vacuum~\cite{schwarz95a,aspinwall95a}!  Then the duality
gives a geometrical explanation of the nonperturbative S duality symmetry of
the IIB theory:  the transformation $\rho \rightarrow - 1/\rho$, which sends
$g_s \rightarrow 1/g_s$ in the IIB theory, corresponds to interchanging the two
cycles of the torus in the M theory description.  To complete the story, we should
relate the area of the M theory torus $(A_M)$ to the radius of the IIB theory
circle $(R_B)$.  This is a simple consequence of formulas
given above
\begin{equation}
m_p^3 A_M = (2 \pi R_B)^{-1}.
\end{equation}
Thus the limit $R_B \rightarrow 0$, at fixed $\rho$, corresponds to
decompactification of the M theory torus, while preserving its shape.
Conversely, the limit $A_M \rightarrow 0$ corresponds to decompactification of
the IIB theory circle.

The duality can be explored further by matching the various $p$-branes in 9
dimensions that can be obtained from either the M theory or the IIB theory
viewpoints~\cite{schwarz95b}.  
When this is done, one finds that everything matches nicely and
that one deduces various relations among tensions, such as
\begin{equation}
T_{M5} = \frac{1}{2\pi} (T_{M2})^2.
\end{equation}
This relation was used earlier when we asserted that $T_{M2} = 2\pi m_p^3$ and
$T_{M5} = 2\pi m_p^6$.

Even more interesting is the fact that the IIB theory contains an infinite
family of strings labelled by a pair of relatively prime integers $(p,q)$~\cite{schwarz95a}.
These integers correspond to string charges that are sources of the gauge
fields $B_{\mu\nu}^{(1)}$ and $B_{\mu\nu}^{(2)}$.  The $(1,0)$ string can be
identified as the fundamental IIB string, while the $(0,1)$ string is the
D-string.  From this viewpoint, a $(p,q)$ string can be regarded as a bound
state of $p$ fundamental strings and $q$ D-strings~\cite{witten95b}.  
These strings have a very
simple interpretation in the dual M theory description.  They correspond to an
M2-brane with one of its cycles wrapped around a $(p,q)$ cycle of the torus.
The minimal length of such a cycle is proportional to $|p+q \tau|$, and thus
(using $\tau = \rho$) one finds that the tension of a $(p,q)$ string is given by
\begin{equation}
T_{p,q} = 2\pi|p + q\rho| m_s^2. \label{pqtension}
\end{equation}
The normalization has been chosen to give
$T_{1,0} = 2\pi m_s^2$. Then (for $\theta = 0$) $T_{0,1} = 2\pi
m_s^2/g_s$, as expected. 
Note that decay is kinematically forbidden by charge conservation
when $p$ and $q$ are relatively prime.  When they have a common division $n$,
the tension is the same as that of an $n$-string system.  Whether or not there
are threshold bound states is a nontrivial dynamical question, which has
different answers in different settings.  In this case there are no such bound
states, which is why $p$ and $q$ should be relatively prime.

Imagine that you lived in the 9-dimensional world that is described
equivalently as M theory compactified on a torus or as the type IIB superstring theory
compactified on a circle.  Suppose, moreover, you had very high energy
accelerators with which you were going to determine the ``true'' dimension of
spacetime.  Would you conclude that 10 or 11 is the correct answer?  If either
$A_M$ or $R_B$ was very large in Planck units there would be a natural choice,
of course.  But how could you decide otherwise?  The answer is that either
viewpoint is equally valid.  What determines which choice you make is which of
the massless fields you regard as ``internal'' components of the metric tensor
and which ones you regards as matter fields.  Fields that are metric components in one
description correspond to matter fields in the dual one.

\section{U Dualities}

Maximal supergravity theories (ones with 32 conserved supercharges) typically
have a noncompact global symmetry group $G$.  For example, in the case of type
IIB supergravity in ten dimensions the group is $SL(2,R)$.  When one does
dimensional reduction one finds larger groups in lower dimensions.  For
example, ${\cal N} = 8$ supergravity in four dimensions 
has a noncompact $E_7$ symmetry~\cite{cremmer78b}.  More
generally, for $D = 11-d$,  $3\leq d \leq 8$, one finds a maximally noncompact
form of $E_d$, denoted $E_{d,d}$.  These are statements about classical field
theory.  The corresponding statement about superstring theory/M theory is that
if we toroidally compactify M theory on $R^D \times T^d$ or type IIB
superstring theory on $R^D \times T^{d-1}$, the resulting moduli space of
theories is invariant under an infinite discrete {\em U duality} group.  The group,
denoted $E_d (Z)$, is a maximal discrete subgroup of the noncompact $E_{d,d}$
symmetry group of the corresponding supergravity theory~\cite{hull94}.  
An example that we
will focus on below is
\begin{equation}
E_3 (Z) = SL(3,Z) \times SL(2,Z).
\end{equation}

The U duality groups are generated by the Weyl subgroup of $E_{d,d}$ plus
discrete shifts of axion-like fields.  The subgroup $SL(d,Z) \subset E_d (Z)$
can be understood as the geometric duality (modular group) of $T^d$ in the M
theory picture.  This generalizes the $SL(2,Z)$ discussed in the preceding
section.  The subgroup $SO (d-1, d-1; Z) \subset E_d (Z)$ is the T duality
group of type IIB superstring theory compactified on $T^{d-1}$.  These two
subgroups intertwine nontrivially to generate the entire $E_d (Z)$ U duality
group.

Suppose we wish to focus on M theory and disregard type IIB superstring theory.
 Then we have a geometric understanding of the $SL(d,Z)$ subgroup of $E_d (Z)$
from considering M theory on $R^{11-d} \times T^d$.  But what does the rest of
$E_d(Z)$ imply?  To address this question it will suffice to consider the first
nontrivial case to which it applies, which is $d=3$.  In this case the U
duality group is $SL(3,Z) \times SL(2,Z)$.  The first factor is geometric from
the M theory viewpoint and nongeometric from the IIB viewpoint, whereas the
second factor is geometric from the IIB viewpoint and nongeometric from the M
theory viewpoint.  So the question boils down to understanding the implication
of the $SL(2,Z)$ duality in the M theory construction.  Specifically, we want to
understand the nontrivial $\tau \rightarrow - 1/\tau$ transformation.

To keep the story as simple as possible, we will take the $T^3$ to be
rectilinear with radii $R_1, R_2, R_3$ ({\it i.e.}, $g_{ij} \sim R_i^2 \delta_{ij}$)
and assume that $C_{123} = 0$.  Let us suppose that $R_3$ corresponds to the
``eleventh'' dimension that takes us to the IIA theory.  Then we have IIA
theory on a torus with radii $R_1$ and $R_2$.  The nongeometric duality 
of M theory is T duality of IIA theory.
T duality gives a mapping to an equivalent point in the moduli space for
which
\begin{equation}
R_i \rightarrow R'_i = \frac{\ell_s^2}{R_i} = \frac{\ell_p^3}{R_3 R_i}\quad  i = 1,2,
\end{equation}
with $\ell_s$ unchanged.  Note that we have used eq.~(\ref{M1}), reexpressed as
$\ell_p^3 = R_3 \ell_s^2$.  Under a T duality the string coupling constant also
transforms.  The rule is that the coupling of the effective theory (8d in this
case) is invariant:
\begin{equation}
\frac{1}{g_8^2} = 4\pi^2 \frac{R_1 R_2}{g_s^2} = 4\pi^2 \frac{R'_1
R'_2}{(g'_s)^2}.
\end{equation}
Thus
\begin{equation}
g'_s = \frac{g_s \ell_s^2}{R_1 R_2}.
\end{equation}

What does this imply for the radius of the eleventh dimension $R_3$?  Using
eq.~(\ref{M2}),
\begin{equation}
R_3 = g_s\ell_s \rightarrow R'_3 = g'_s \ell_s.
\end{equation}
Thus
\begin{equation}
R'_3 = \frac{g_s\ell_s^3}{R_1R_2} = \frac{\ell_p^3}{R_1 R_2}.
\end{equation}
However, the 11d Planck length also transforms, because
\begin{equation}
\ell_p^3 = g_s \ell_s^3 \rightarrow (\ell'_p)^3 = g'_s \ell_s^3
\end{equation}
implies that
\begin{equation}
(\ell'_p)^3 = \frac{g_s\ell_s^5}{R_1 R_2} = \frac{\ell_p^6}{R_1R_2 R_3}.
\end{equation}

The perturbative IIA description is only applicable for $R_3 \ll R_1, R_2$.
However, even though T duality was originally discovered in perturbation
theory, it is supposed to be an exact nonperturbative property.  Therefore this
duality mapping should be valid as an exact symmetry of M theory without any
restriction on the radii.  Another duality is an interchange of circles, such
as $R_3 \leftrightarrow R_1$.  This corresponds to the nonperturbative S
duality of the IIB theory, as we discussed earlier.  Combining these dualities
we obtain the desired nongeometric duality of M theory on $T^3$~\cite{blau97}.  
It is given by
\begin{equation}\label{a}
R_1 \rightarrow \frac{\ell_p^3}{R_2R_3},
\end{equation}
and cyclic permutations, accompanied by
\begin{equation}\label{b}
\ell_p^3 \rightarrow \frac{\ell_p^6}{R_1 R_2 R_3}.
\end{equation}

Equations~(\ref{a}) and~(\ref{b}) have a nice interpretation.
Equation~(\ref{a}) implies that 
\begin{equation}
{1 \over R_1} \rightarrow (2\pi R_2) (2\pi R_3) T_{M2} .
\end{equation}
Thus it interchanges Kaluza--Klein excitations with wrapped supermembrane
excitations.  It follows that these six 0-branes belong to the ({\bf 3}, {\bf 2})
representation of the U-duality group.
Equation~(\ref{b}) implies that
\begin{equation}
T_{M2} \rightarrow (2\pi R_1) (2\pi R_2) (2\pi R_3) T_{M5}.
\end{equation}
Therefore it interchanges an unwrapped M2-brane with an M5-brane wrapped on the
$T^3$.  Thus these two 2-branes belong to the ({\bf 1}, {\bf 2})
representation of the U-duality group.
This basic nongeometric duality of M theory, combined with the geometric
ones, generates the entire U duality group in every dimension.  It is a
property of quantum M theory that goes beyond what can be understood from the
effective 11d supergravity, which is geometrical.

This analysis has been extended to allow 
$C_{123} \not= 0$~\cite{obers}.  In
this case there are indications that the torus should be considered to be
{\em noncommutative}~\cite{douglas97}.

\section{The D3-Brane and ${\cal N}=4$ Gauge Theory}

D-branes have a number of special properties, which make them especially
interesting.   By definition, they are branes on which strings can end---D
stands for {\em Dirichlet} boundary conditions.  The end of a string carries a
charge, and the D-brane world-volume theory contains a $U(1)$ gauge field that carries the
associated flux.  When $n$ D$p$-branes are coincident, or parallel and nearly
coincident, the associated $(p + 1)$-dimensional world-volume theory is a
$U(n)$ gauge theory.  The $n^2$ gauge bosons $A_\mu^{ij}$ and their
supersymmetry partners arise as the ground states of oriented strings running
from the $i$th D$p$-brane to the $j$th D$p$-brane.  The diagonal elements,
belonging to the Cartan subalgebra, are massless.  The field
$A_\mu^{ij}$ with $i \not= j$ has a mass proportional to the separation of the
$i$th and $j$th branes.  This separation is described by the vev of a
corresponding scalar field in the world-volume theory.

The $U(n)$ gauge theory associated with a stack of $n$ D$p$-branes has maximal
supersymmetry (16 supercharges).  The low-energy effective theory, when the
brane separations are small compared to the string scale, is supersymmetric
Yang--Mills theory.  These theories can be constructed by dimensional reduction
of 10d supersymmetric $U(n)$ gauge theory to $p+1$ dimensions.  In fact,
that is how they originally were constructed~\cite{brink77}.  For $p\leq 3$, the low-energy
effective theory is renormalizable and defines a consistent quantum theory.
For $ p = 4,5$ there is good evidence for the existence nongravitational
quantum theories that reduce to the gauge theory in the infrared.  For $p\geq
6$, it appears that there is no decoupled nongravitational quantum theory~\cite{sen97}.

A case of particular interest, which we shall now focus on, is $p = 3$.  A stack of
$n$ D3-branes in type IIB superstring theory has a decoupled ${\cal N} = 4, $ $d = 4$
$U(n)$ gauge theory associated to it.  This gauge theory has a number of special
features.  For one thing, due to boson--fermion cancellations, there are no
$UV$ divergences at any order of perturbation theory.  The beta function
$\beta(g)$ is identically zero, which implies that the theory is scale
invariant (aside from scales introduced by vevs of the scalar fields).  In
fact, ${\cal N}=4, $ $d=4$ gauge theories are conformally invariant.  The conformal
invariance combines with the supersymmetry to give a superconformal symmetry,
which contains 32 fermionic generators.  Half are the ordinary linearly
realized supersymmetrics, and half are nonlinearly realized ones associated to
the conformal symmetry.  The name of the superconformal group in this case is
$SU(4|4)$.  Another important property of ${\cal N}=4$, $ d=4$ gauge theories is
electric-magnetic duality~\cite{montonen77}.  
This extends to an $SL(2,Z)$ group of dualities.
To understand these it is necessary to include a vacuum angle $\theta_{YM}$ and
define a complex coupling
\begin{equation}
\tau = \frac{\theta_{YM}}{2\pi} + i \frac{4\pi}{g_{YM}^2}.
\end{equation}
Under $SL(2,Z)$ transformations this coupling transforms in the usual nonlinear
fashion $\left(\tau \rightarrow \frac{a\tau+b}{c\tau+d}\right)$ and the
electric and magnetic fields transform as a doublet.  Note that the conformal
invariance ensures that $\tau$ is a meaningful scale-independent constant.

Now consider the ${\cal N}=4 $ $U(n)$ gauge theory associated to a stack of $n$
D3-branes in type IIB superstring theory.  
There is an obvious identification, that turns out to be correct.
Namely, the $SL(2,Z)$ duality of the gauge theory is induced from that of the
ambient type IIB superstring theory.  In particular, the $\tau$ parameter of the
gauge theory is the value of the complex scalar field $\rho$ of the string
theory.  This makes sense because $\rho$ is constant in the field configuration
associated to a stack of D3-branes.
The D3-branes themselves are invariant under $SL(2,Z)$ transformations.  Only
the parameter $\tau = \rho$ changes, but it is transformed to an equivalent
value.  All other fields, such as $B_{\mu\nu}^{(i)}$, which are not invariant,
vanish in this case.

As we have said, a fundamental $(1,0)$ string can end on a D3-brane.  But by
applying a suitable $SL(2,Z)$ transformation, this configuration is transformed
to one in which a $(p,q)$ string---with $p$ and $q$ relatively prime---ends on
the D3-brane.  The charge on the end of this string describes a dyon with
electric charge $p$ and magnetic $q$, with respect to the appropriate gauge field.  
More generally, for a stack of $n$
D3-branes, any pair can be connected by a $(p,q)$ string.  The mass is
proportional to the length of the string times its tension, which we saw is
proportional to $|p + q\rho|$.  In this way one sees that the electrically
charged particles, described by fundamental fields, belong to infinite
$SL(2,Z)$ multiplets.  The other states are nonperturbative excitations of the
gauge theory.  The field configurations that describe them
preserve half of the supersymmetry.  As a
result their masses saturate a BPS bound and are given exactly by the
considerations described above.

An interesting question, whose answer was unknown until recently, is whether
${\cal N}=4 $  gauge theories in four dimensions
also admit nonperturbative excitations that preserve
1/4 of the supersymmetry.  To explain the answer, it is necessary to first make
a digression to consider three-string junctions.

As we have seen, type IIB superstring theory contains an infinite multiplet of
strings labelled by a pair of relatively prime integers $(p,q)$.  Three
strings, with charges $(p_i, q_i), $ $i = 1,2,3,$ can meet at a point
provided that charge is conserved~\cite{aharony96,schwarz96}.  This means that
\begin{equation}\label{x}
\sum p_i = \sum q_i = 0,
\end{equation}
if the three strings are all oriented inwards.  
(This is like momentum conservation in an ordinary Feynman diagram.) Such a
configuration is stable, and preserves 1/4 of the ambient supersymmetry
provided that the tensions balance.  It is easy to see how this can be
achieved.  If one regards the plane of the junction as a complex plane and
orients the direction of a $(p,q)$ string by the phase of $p + q\tau$, then
eqs.~(\ref{pqtension}) and (\ref{x}) ensure a force balance.

The three-string junction has an interesting dual M theory interpretation.  If
one of the directions perpendicular to the plane of the junction is taken to be
a circle, then we have a string junction in nine dimensions.  This must have a
dual interpretation in terms of M theory compactified on a torus.  We have
already seen that a $(p,q)$ string corresponds to an M2-brane with one of its
cycles wrapped on a $(p,q)$ cycle of the torus.  So now we join three such
cylindrical membranes together.  Altogether we have a single smooth M2-brane
forming a $Y$, like a junction of pipes.  
The three arms are wrapped on
$(p_i, q_i)$ cycles of the torus.  This is only possible topologically
when eq.~(\ref{x}) is satisfied.

We can now describe a pretty construction of 1/4 BPS states 
in ${\cal N}=4$ gauge theory, due to Bergman~\cite{bergman97}.
Such a state is described by a 3-string junction, with the three prongs
terminating on three different D3-branes.  This is only possible for $n \geq
3$, which is a necessary condition for 1/4 BPS states.  The mass of such a
state is given by summing the lengths of each string segment weighted by its
tension.  This gives a result in agreement with the BPS formula.
Clearly this is just the beginning of a long story, since the simple
picture we have described can be generalized to arbitrarily complicated string
webs.  So long as the web is in a plane, charges are conserved at the
junctions, and all string segments are oriented in the way we have described,
the configuration will be 1/4 BPS.  Remarkably, arbitrarily high spins can
occur.  There are simple rules for determining them~\cite{bergman98}.  
When the web is
nonplanar, supersymmetry is completely broken, and reliable mass
calculations become difficult.  However, one should still be able to achieve a
reliable qualitative understanding of such excitations.  In general, there are
regions of moduli space in which such nonsupersymmetric states are stable.

\section{Conclusion}

In this brief review we have described some of the interesting advances
in understanding superstring theory that have taken place in the past
few years. Many others, such as studies of black hole entropy, have not even been
mentioned. The emphasis has been on the nonperturbative appearance
of an eleventh dimension in type IIA superstring theory, as well as
its implications when combined with superstring T dualities. In particular,
we argued that there should be a consistent quantum vacuum, whose
low-energy effective description is given by 11d supergravity. 
The relevant quantum theory -- called M theory -- has important
features, such as the nongeometric U duality described in section 4, that go beyond 
what can be understood within ordinary (nonrenormalizable) 11d supergravity. 

What we have described makes a convincing self-consistent picture, but it
does not constitute a complete formulation of M theory. In the past two years
there have been some major advances in that direction, which we will 
briefly mention here. The first, which goes by the name 
of {\em Matrix Theory}~\cite{banks96},
bases a formulation of M theory in flat 11d spacetime in terms of the
supersymmetric quantum mechanics of N D0-branes in the large N limit.
This proposal has been generalized to include an interpretation for finite
N.  In that case Susskind has proposed an identification with {\em discrete
light-cone quantization} of M theory, in which there are N units of momentum
along a null compact direction~\cite{susskind97}. 
Both versions of  Matrix Theory have passed
all tests that have been carried out, some of which are very nontrivial. At
times there appeared to be discrepancies, but these were all the result of
subtle errors that have now been tracked down. The construction has a nice
generalization to describe compactification of M theory 
on a torus $T^n$~\cite{taylor96}. 
However, it does not seem to be useful for $n > 5$~\cite{sen97}, and other
compactification manifolds are (at best) awkward to handle. Another
shortcoming of this approach is that it treats the eleventh dimension
differently from the other ones. 

Another proposal relating superstring and M theory backgrounds to large N
limits of certain field theories has been put forward 
recently by Maldacena~\cite{maldacena97} and made 
more precise by others~\cite{gubser98}.
In this approach, there is a conjectured duality ({\it i.e.}, equivalence) between a
conformally invariant field theory (CFT) in $n$ dimensions and type IIB superstring theory
or M theory on an Anti-de-Sitter space (AdS) in $n+1$ dimensions. The remaining
$9-n$ or $10-n$ dimensions form a compact space, the simplest cases being spheres.
The three examples with unbroken supersymmetry are $AdS_5 \times S^5$,
$AdS_4 \times S^7$, and $AdS_7 \times S^4$.
This approach is sometimes referred to as {\em AdS/CFT duality}.
This is an extremely active and very promising subject. It has already taught us
a great deal about the large N behavior of various gauge theories. 
As usual, the easiest theories to study are ones with a lot of supersymmetry, 
but it appears that in this approach supersymmetry breaking is more accessible than
in previous ones. For example, it might someday be possible to construct the QCD
string in terms of a dual AdS gravity theory, and use it
to carry out numerical calculations of the hadron spectrum. 
Indeed, there have already been some 
preliminary steps in this direction~\cite{ooguri98}.

Despite all of the successes that have been achieved in advancing our understanding of
superstring theory and M theory, there clearly is still a long way
to go.  In particular, despite much effort and several imaginative proposals,
we still do not have a convincing mechanism for ensuring the vanishing
(or extreme smallness) of the cosmological constant for nonsupersymmetric vacua.
Superstring theory is a field with very ambitious goals. The remarkable fact is
that they still seem to be realistic. However,
it may take a few more revolutions before they are attained.

\end{document}